\documentclass[
10pt,twocolumn,notitlepage,oneside]{revtex4}
\usepackage{rotating}
\usepackage{amsfonts,longtable}
\usepackage{amsmath}
\renewcommand{\baselinestretch}{1}
\def\MspcII{$\textrm{M}_{\odot}/\textrm{pc}^{2}$}

\def\kms{$\textrm{km/s}$}
\def\Mpc{$\textrm{Mpc}$}
\def\kpc{kpc}

\def\H2{H$_{2}$}
\def\HI{HI}
\def\roH2{$\rho_{\textrm{H}_2}$}

\def\MH2{M$_{\textrm{H}_2}$}

\begin{document}

\title{Atomic and Molecular Gas Components in Spiral Galaxies \\
of the Virgo Cluster}
\author{\firstname{A.~V.}~\surname{Kasparova}}
\email[]{anastasya.kasparova@gmail.com}
\affiliation{Sternberg Astronomical Institut M. V. Lomonosov Moscow
State Univ. }
\begin{abstract}
Based on two models, we investigate the molecular-to-atomic gas
ratio in Virgo cluster galaxies in comparison with field galaxies.
We show that the enhanced metallicity for cluster members and the
ram pressure stripping of atomic gas from the disk periphery cannot
fully explain the observed gas component ratios. The additional
environmental factors affecting the interstellar medium and leading
to an increase in the molecular gas fraction should be taken into
account for cluster galaxies.


\vspace{0.2cm}

\textit{Astronomy Letters, 2012, Vol. 38, No. 2, pp. 63--73.}

\end{abstract}
\maketitle
\section{INTRODUCTION}

Atomic and molecular hydrogen are the main components of the
interstellar medium, with the mechanisms of the transition from one
gas state to the other being known poorly and being the subject of
an active discussion. Stars are generally believed to be formed in
dense molecular clouds and the transition from warm atomic hydrogen
to cold molecular one is the most important step in this process,
although the formation of stars directly from {\HI} is also
considered (Glover and Clark 2011). Nevertheless, the fact that the
distribution of young stars in galaxies correlates precisely with
\H2 regions is beyond question (Leroy et al. 2008; Bigiel et al.
2011). Since direct observations of molecular clouds cause many
difficulties, the cold \H2 gas is observed predominantly by
accompanying molecules, primarily CO (the J = 1.0 transition at 115
GHz), which take the second place in abundance in the interstellar
medium after \H2.

Determining the conversion factor $\chi$ between the CO intensity
and the molecular hydrogen surface density is a separate problem.
Observations of our Galaxy and its immediate neighborhood (see
Bolatto et al. 2008; Leroy et al. 2011), where the molecular gas
regions can be studied in more detail, play a crucial role here.
There is reason to believe that $\chi$ is constant in a fairly wide
range of gas densities— from diffuse \H2 to dense self-gravitating
clouds (Liszt et al. 2010). However, it depends on metallicity, at
least in the case of a strong heavy elements deficiency (Boselli et
al. 2002; Leroy et al. 2011).

\renewcommand{\baselinestretch}{1}
\begin{table*}

\begin{tabular}{lccccccccccl}
\hline \hline
 &NGC & Messier & $D_{M87}$&$R_{25}$&type& $V_r$ &\multicolumn{2}{c}{deficiency}&\multicolumn{2}{c}{references}&N\\
 &    &         & deg  & arcmin  &   &km/s&\H2&\HI&\H2&\HI&\\
\hline%
(1)&(2)& (3)    & (4)   & (5)  & (6)  & (7)  &(8)&(9)&(10)&(11)&(12)\\
\hline%
1&4254& 99 & 3.6 &2.69& 5.2&2410&-0.20&0.18&[1]&[2]&I\\%
2&4298&     &3.2&1.48& 5.2&1140&  && [1]&[3]&III\\%
3&4302&     &3.1&2.45& 5.4&1118&  && [1]&[3]&III\\%
4&4303&61  &8.2&3.46& 4.0&1570&-0.08&0.13&[1]&[2]&I\\%
5&4321& 100&3.9&3.01& 4.1&1579&0.05&0.53&[1]&[2]&I\\%
6&4402&     &1.4&1.77& 3.2&119 &0.02&0.73&[1]&[2]&III\\%
7&4419&     &2.8&1.95& 1.1& -254&0.40&1.04&[1]&[3]&II, III\\%
8&4501&88  &2.1&3.38& 3.4& 2280&0.14&0.51&[1]&[2]&I\\%
9&4535&     &4.3&3.46& 5.0& 1958&0.02&0.35&[1]&[2]&II\\%
10&4536&     &10.2&3.54& 4.3& 1807&0.18&0.29&[1]&[3]&I\\%
11&4548&91  &2.4&2.63& 3.1& 486&0.84&0.70&[1]&[2]&II\\%
12&4567&     &1.8&1.38& 4.0& 2265&  &&[1]&[3]&I\\%
13&4568&     &1.8&1.19& 4.1& 2255&  &&[1]&[3]&I\\%
14&4569&90  &1.7&4.56& 2.4& -233&0.23&1.13&[1]&[2]&I\\%
15&4579&58  &1.8&2.51& 2.8& 1518&0.62&0.71&[1]&[2]&II\\%
16&4647&     &3.2&1.41& 5.2& 1415&  &&[1]&[2]&I\\%
17&4654&     &3.3&2.51& 5.9& 1034&-0.02&0.17&[1]&[2]&I\\%
18&4689&     &4.3&2.29& 4.7& 1613& 0.14&1.06&[1]&[2]&I\\%
\hline \hline
\end{tabular}
\caption{Parameters of the Virgo cluster galaxies. Columns 8 and 9
give the references to the data on the molecular and atomic hydrogen
profiles that we used: [1]---Chung et al. (2009a), [2]---Cayatte et
al. (1994), [3]---Chung et al. (2009b).} \label{tab1}
\end{table*}

The radial distributions of the atomic and molecular gases differ
greatly: whereas the azimuthally averaged {\HI} surface density over
a large extent changes only slightly along the radius and is
$\sim10$~{\MspcII} for normal galaxies for the inner disk regions,
the CO emission is more strongly concentrated to the center.
Investigating the \H2/\HI{} ratio and searching for the key
parameters of the interstellar medium responsible for the balance
between these components, which is of crucial importance for
understanding the mechanisms inducing the formation of stars and
inhibiting this process, are of particular interest.

There are exist several theoretical models for the formation of
molecular clouds (Elmegreen 1989; McKee and Krumholz 2010;
Girichidis et al. 2011). For a wide range of parameters, these
models are in good agreement with the well-known empirical
relationship between the gas surface density ratio
$\eta=\Sigma_{H_2}/\Sigma_{HI}$ and the equilibrium turbulent
interstellar gas pressure $P$ changing along the radii of galaxies
(Blitz and Rosolowsky 2006). The relationship exists despite the
fact that the \H2 distribution in galaxies has a complex structure:
part of \H2 is in a diffuse form and part is in giant molecular
clouds, with the ratio of these components being not quite certain
and may change with galactocentric distance (Rosolowsky and Blitz
2005). In the regions where the molecular gas in the form of
self-gravitating clouds dominates, the dependence $\eta(P)$ is, in
general, not obvious. The idea that the balance between {\HI} and
\H2 depends not so much on the pressure as on the total surface
density and metallicity of the interstellar gas (Krumholz et al.
2009) competes with the latter model. The amount of heavy elements
controls the formation of \H2 molecules, while the layer of
interstellar gas shields them from the destructive action of
ultraviolet radiation.

For normal galaxies, the dependences $\eta(P)$ and
$\eta(\Sigma_{gas})$ show an equally good correlation (Krumholz et
al. 2009) and it is rather difficult to determine which of the
physical mechanisms is decisive. Nevertheless, under special
conditions, for example, in metal-poor dwarf galaxies, it becomes
possible to check the validity of these models (Fumagalli et al.
2010). For such objects, the discrepancy between the dependences
being discussed is significant, and the correlation
$\eta(\Sigma_{gas})$ probably better corresponds to the observed
characteristics of the gas medium than $\eta(P)$. Nevertheless, the
cloud formation models were calculated by Krumholz et al. (2009) by
taking into account the complex mass transfer processes in the
medium but under a large number simplifications for a fairly
simplegeometry and without allowance for the possible environmental
effects.

In this paper, we focus our attention on spiral galaxies of the
Virgo cluster distinguished by a great variety of relative \H2
abundances. Checking the relations $\eta(P)$ and
$\eta(\Sigma_{gas})$ for these objects will show within the
framework of which model the environmental effects on the
interstellar medium can be described more properly and what
additional physical parameters should be included in the
description. Cluster galaxies are subjected both to the
gravitational effect from other cluster members and to the influence
of the intergalactic medium. As a result, a deficiency of atomic
hydrogen is formed in galactic disks (Vollmer et al. 2001; Roediger
2009). However, the changes in {\HI} abundance generally have no
effect on the total \H2 mass (Kenney and Young 1986, 1989), and only
in rare cases does the environmental effects lead to a strong
deficiency of both components (Fumagalli et al. 2009). In this
paper, we will concentrate on studying the properties of the atomic
and molecular interstellar medium in Virgo cluster galaxies. In our
subsequent publications, we are planning to investigate in more
detail the possible causes of the peculiarities of the relationship
between these components.

The paper consists of three main sections: in Section 2, we discuss
the technique used to numerically calculate the hydrostatic gas
pressure in the disk midplane and present the results for the
galaxies of our sample; in Section 3, we consider in detail the
characteristics of the gas components of these objects; in Section
4, we discuss our results and the prospects for further studies.

\section{ESTIMATING THE GAS PRESSURE IN~THE DISK MIDPLANE}

Blitz and Rosolowsky (2006) calculated the equilibrium gas pressure
P in the disk midplane under considerable simplifications and
without allowance for the gas self-gravity and the dark halo effect,
which become significant on the periphery of galactic disks.
However, precisely the outer regions of the gas disks affected most
strongly by the environment are most interesting for our purposes.
Therefore, here we use a more accurate self-consistent method and
apply it to two samples: a sample of Virgo cluster galaxies and, for
comparison, a sample of field galaxies. To derive the radial
pressure profiles, we use the observed {\HI} and \H2 surface
densities, stellar surface brightnesses, integrated color indices,
and rotation curves for determining the stellar disk velocity
dispersion.

\subsection {The Pressure Estimation Technique}

\begin{table}[h]
\begin{tabular}{lccccccc}
\hline \hline
 &NGC & Messier & $D$ &$R_{25}$&type&\multicolumn{2}{c}{deficiency}\\
 &    &         & Mpc & kpc  &      &\H2&\HI\\
\hline%
(1)&(2)& (3)    & (4)   & (5)  & (6)  & (7)  &(8)\\
\hline%
1&628 &  74     &7.3  &10.4  &5.2&0.19&-0.01\\%
2&2841&         &14.1 &14.2  &3.0&&\\
3&3184&         &11.1 &11.9  &6.0&0.11&0.32\\%
4&3198&         &13.8 &13.0  &5.2&&\\%
5&3351&  95     &10.1 &10.6  &3.1&0.34&0.60\\
6&3521&         &10.7 &12.9  &4.0&0.19&0.06\\%
7&3627&  66     &9.3  &13.9  &3.1&-0.11&0.81\\
8&4736&  94     &4.3  &5.3   &2.4&0.78&0.68\\%
9&5055&  63     &10.1 &17.4  &4.0&0.04&0.12\\%
10&5194& 51     &8.4  &27.4  &4.0&-0.36&0.12\\%
11&6946&        &5.9  &9.8   &5.9&-0.28&-0.45\\%
12&7331&        &14.7 &19.6  &3.9&0.14&-0.04\\%

\hline \hline
\end{tabular}
\caption{Parameters of the field galaxies.} \label{tab2}
\end{table}

In present-day works, the Poisson and hydrostatic equilibrium
equations are simultaneously solved to estimate the gas pressure in
the disk midplane. In this case, a number of simplifying assumptions
are usually made. For example, Blitz and Rosolowsky (2004, 2006)
calculated the pressure for infinite two-component disks (stars and
gas) with a vertical scale height of the gas layer much smaller than
that of the stellar disk and without allowance for the spheroidal
components and gas self-gravity. Under these assumptions, they
obtained an expression for the gas pressure:

\begin{equation}
P =
0.84(G\Sigma_{star})^{0.5}\Sigma_{gas}\frac{v_{gas}}{h_{star}^{0.5}},
\label{Press2}
\end{equation}
where $G$ is the gravitational constant, $\Sigma_{star}$ and
$\Sigma_{gas}$ are the stellar and gas surface densities, $v_{gas}$
is the turbulent gas velocity dispersion, and $h_{star}$ is the
stellar scale height. The authors assumed that the last two
parameters changed only slightly both with distance from the
object's center and from galaxy to galaxy. In this approach, the
pressure turns out to be a function of only the stellar and gas
surface densities. However, as was justly pointed out by the authors
themselves, when the gas becomes predominantly molecular (in the
central regions of galaxies), the linear dependence of
$\eta=\Sigma_{H_2}/\Sigma_{HI}$ on the pressure reflects the method
of estimating $P$ than the physical relationship between the
quantities being compared. As was shown by Kasparova and Zasov
(2008), the simplifications listed above for an individual galaxy
lead to a considerable underestimation of the pressure (up to 40\%)
at distances greater than or of the order of $R_{25}$, where the gas
self-gravity and the presence of a dark halo can play a significant
role. However, for our goals, the disk peripheries are of the
primary interest, because the influence of the intergalactic medium
is most significant precisely on them. Therefore, here we use a more
appropriate method of calculating the turbulent gas pressure.

We calculate the gas pressure from the gas volume density in the
disk midplane under the assumption of a radius-independent turbulent
velocity $v_{gas}$ but with allowance made for the gas self-gravity,
the change in stellar disk thickness with radius, and the dark halo
contribution to the galactic gravitational potential. The volume
density was found through a self-consistent solution of the
equations that describe the vertical structure of the stellar,
atomic, and molecular disk components. This approach was proposed by
Narayan and Jog (2002) for our Galaxy and was developed by Kasparova
and Zasov (2008) and Abramova and Zasov (2011). Since the turbulent
gas velocity does not change greatly from galaxy to galaxy, we take
it to be 9 and 6 {\kms} (in one coordinate) for the atomic and
molecular gases, respectively. Note that the result is barely
sensitive to the choice of $v_{gas}$: if we underestimated this
velocity, then this would lead to an overestimation of the resultant
volume density of the component that would largely ``quench'' the
effect of the error in the gas velocity dispersion on the resultant
pressure (a twofold increase in $v_{gas}$ will cause $P$ to rise
only by a factor of 1.5 rather than 2, as follows from the
simplified formula (1)).

The final expression for each galactic disk component derived from
the Poisson and hydrostatic equilibrium equations is

\begin{equation}
\frac{d^2\rho_i}{dz^2} =
\frac{\rho_i}{\langle(v_z)^2_i\rangle}\left[-4\pi
G\sum_{i=1}^3\rho_i-\frac{\partial^2\phi_d}{\partial z^2}\right]+
\frac{1}{\rho_i}\left(\frac{d\rho_i}{dz}\right)^2, \label{eq12}
\end{equation}
where the term in square brackets corresponds to the potential of a
three-component axisymmetric disk inside a halo, the index $i$
denotes one of the disk components—the stars, \HI, or \H2, while
$\phi_d$ corresponds to the potential of the dark halo that we
assumed to be pseudo-isothermal. The system of equations for the
stellar disk and the gas subsystems was solved numerically by the
fourth-order Runge--Kutta method with boundary conditions in the
disk midplane $z = 0$: $\rho_i = (\rho_0)_i$ and $d\rho_i/dz = 0$
(for more details, see Kasparova and Zasov 2008). We determined the
stellar velocity dispersion from the rotation curve and the surface
density of the disk under the assumption of its marginal stability.
Note that for this method of calculation, the resultant stellar
volume density, to a first approximation, depends only on the local
epicyclic frequency but not on the surface density (see the Appendix
in Abramova and Zasov (2011)). Another, the most popular method of
estimating the stellar density, when the stellar disk thickness is
constant (assuming the vertical and horizontal scale lengths of the
stellar disks to be proportional), gives volume densities of the
components that agree well with the previous method (Kasparova and
Zasov 2008; Abramova and Zasov 2011). As a result, for a number of
galaxies that both belong and do not belong to the Virgo cluster, we
obtained radial volume density distributions for the gas components
in the disk midplane proportional to the equilibrium turbulent
pressure.

It is worth noting that using the azimuthal averaging in our
calculations can introduce an uncertainty, because it is unknown how
strongly the cluster environment affects the homogeneity of the gas
components in azimuthal angle in the central disk regions. In
prospect, it will not be out of place to investigate the influence
of the cluster environment using two-dimensional distributions.

\subsection{Dependence of the Molecular Gas\\ Fraction on the Gas Pressure}

To ascertain how the membership of a galaxy in a cluster affects the
relationship between the molecular gas fraction and pressure, we
calculated the radial gas pressure profiles for 18 Virgo cluster
galaxies and, for comparison, 12 well-studied nearby spirals outside
clusters (field galaxies from here on). The objects were chosen so
that the observational data for them were as homogeneous as
possible.

In Table 1, columns 2 and 3 give the NGC and Messier numbers of the
Virgo cluster galaxies; column 4---the projected distance to the
Virgo cluster center (in degrees); column 5---the angular optical
radius of the galaxy $R_{25}$; column 6---the Hyper Leda
morphological type; column 7---the heliocentric velocity; columns 8
and 9---the references to the sources of \H2 and {\HI} data used:
[1]---Chung et al. (2009a), [2]---Cayatte et al. (1994), [3]---Chung
et al. (2009b); column 10---the numbers of the subsamples to which
we assigned the objects (they will be discussed below). We assume
the distances to all cluster galaxies to be 16.1 {\Mpc} (Ferrarese
et al. 1996). Table 2 provides data on the field galaxies.

\begin{figure*}
\includegraphics[width=16.0cm]{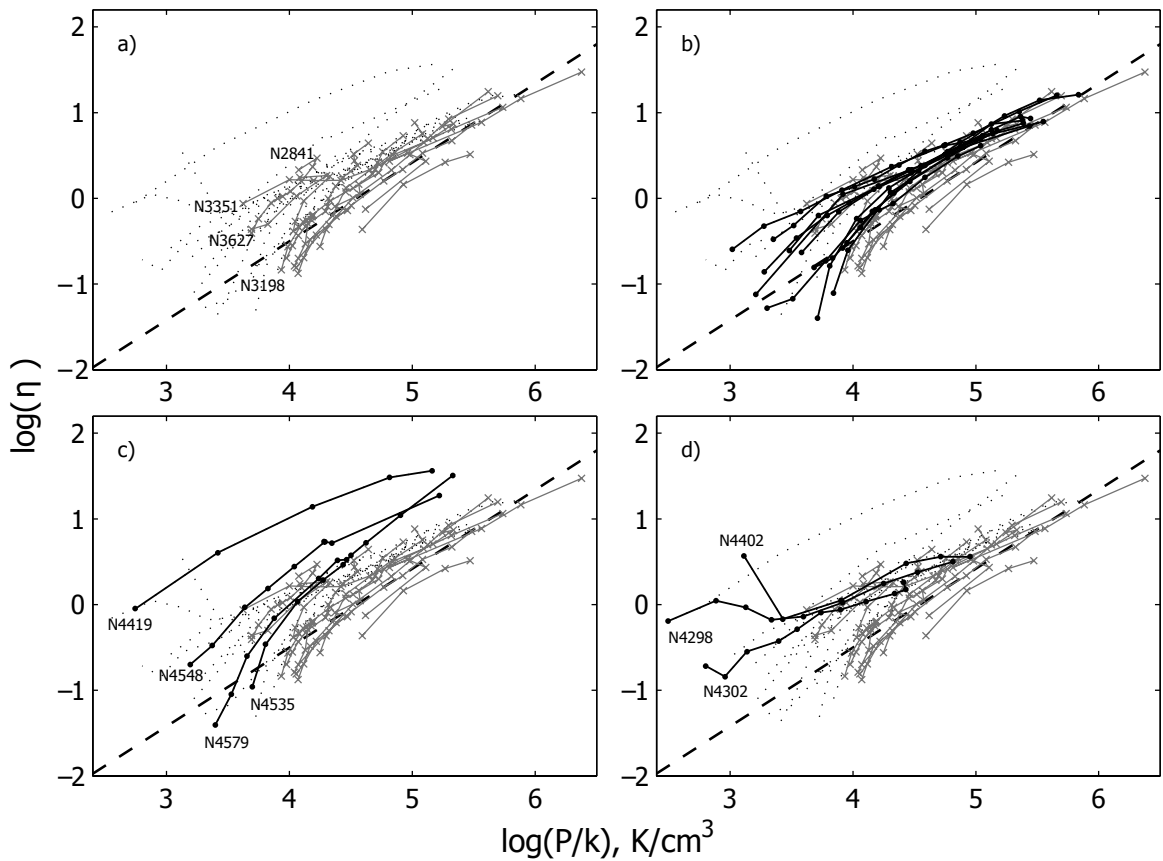}
\caption{Molecular hydrogen fraction $\eta$ versus turbulent gas
pressure in the disk midplane for the field (a) and cluster (b, c,
d) galaxies. On all plots, the gray lines with crosses indicate the
field galaxies and the dotted lines indicate the Virgo cluster
galaxies. The dashed straight line corresponds to the dependence
$\eta\propto P^{0.92}$ (Blitz and Rosolowsky 2006). The cluster
galaxies were divided into three groups described in the text. For
clarity, each group is displayed on separate plots in comparison
with the field galaxies: (b) for the Virgo cluster galaxies
belonging to group I, (c) for group II, and (d) for group
III.}\label{fig1}
\end{figure*}

To calculate the gas pressure, we used the radial distributions of
the azimuthally averaged atomic and molecular hydrogen surface
densities. For the cluster galaxies, we used the catalog of CO
observations (Chung et al. 2009a) with a 45'' resolution and the
{\HI} catalog by Cayatte et al. (1994) with the same resolution. For
several objects absent in Cayatte et al. (1994), we took the data
from Chung et al. (2009b) reduced to the same resolution. For the
field galaxies, the data on the gas components were taken from Leroy
et al. (2008). For all galaxies, the radial \H2 profile was taken
from the published data on $^{12}CO(J=1-0)$ using the conversion
factor $\chi=N_{H_2}/I_{CO}=2\times10^{20}$ $(K\cdot
km/s)^{-1}cm^{-2}$. We also used the observed rotation curves and
the surface brightnesses of the stellar disks from Koopmann et al.
(2001), Dicaire et al. (2008), Rubin et al. (1999), Heald et al.
(2007), Sofue and Rubin (2001), Gavazzi et al. (2003), Nishiyama et
al. (2001), Daigle et al. (2006) and Devereux et al. (1992). These
were converted to the stellar surface density via the mass-to-light
ratio corresponding to the integrated color index (Bell and de Jong
2001) in the infrared from NED. The central regions at $R<1$ {\kpc}
were not considered primarily because of the bulge effect
disregarded in our models.

We plotted the molecular gas fraction $\eta$ against the total
hydrostatic gas pressure $P=\rho_{HI} v_{HI}^2+\rho_{H_2} v_{H_2}^2$
(in the approximation of a continuous medium) in the disk midplane
(Fig. 1) for all of our galaxies.

In Fig. 1, the large scatter of curves relative to the
Blitz--Rosolowsky linear dependence engages our attention. In
contrast to most of the field galaxies, a significant fraction of
the cluster members at the same gas pressure have a higher \H2
fraction, especially on the periphery, than that following from the
Blitz--Rosolowsky dependence. It may be caused by the fact that
either we grossly underestimate the pressure acting on the
interstellar medium and contributing to the phase transition to
molecules or the cluster galaxies are characterized by different
conditions determining the content of the molecular medium. It is
worth noting that some of the field galaxies (NGC 2841, NGC 3351,
and NGC 3627) also lie in the region of an enhanced molecular gas
fraction. This probably reflects the peculiarities of their
evolution, which requires a separate consideration.

The Virgo cluster galaxies can be arbitrarily divided into three
subsamples with a different behavior of the dependence $\eta(P)$
(see the last column in Table 1). For clarity, each group in Fig. 1
is highlighted by the thick lines.

The first group (Fig. 1b) includes the galaxies whose behavior in
the central regions is almost indistinguishable from that of the
field galaxies, but there is a significant dispersion on the
periphery and the molecular gas fraction is higher on average than
that expected at a given pressure (or a pressure that is an order of
magnitude lower at a given $\eta$). Among the objects of the first
group, there is a close pair, NGC 4567/NGC 4568, and this suggests
that the observed scatter on the dependence $\eta(P)$ is not related
to the local interaction between close galaxies.

We attribute the objects for which the molecular gas fraction is
higher than that for single galaxies even in the central regions to
the second group. It includes three galaxies: NGC 4419, NGC 4535,
and NGC 4548. The dependences for them (Fig. 1c) lies much higher
than for the field galaxies. Besides, there is reason to suggest
that the galaxy NGC 4579 also belongs to this group, although,
according to the main source of {\HI} observations that we use (for
data homogeneity) (Cayatte et al. 1994), the profile in the central
regions was not measured. Nevertheless, according to more recent
data (Chung et al. 2009b), a dip is noticeable at the center of the
$\Sigma_{HI}(R)$ distribution.

We place the galaxies NGC 4298, NGC 4402, and NGC 4302 in the third
group. They are characterized by a considerable excess of molecular
hydrogen (or a deficiency of atomic one) for a given pressure on the
disk periphery—in fact, the anticorrelation region (Fig. 1d).
Formally, we can also point out the galaxy NGC 4419 entering the
second group by its anomalously high value of $\eta$ on the
periphery. For three of these objects, $\eta$ is an order of
magnitude higher than that typical of the field galaxies.

It is worth noting that the large scatter on the dependence
$\eta(P)$ for the Virgo cluster galaxies cannot be explained by the
pressure underestimation due to the inaccuracy of the approximate
model. Besides, we have no reasons to suggest that our model is
unsuitable precisely for the cluster galaxies, although we assumed
the disks to be coplanar and axisymmetric, while the axial symmetry
of the gas components of the cluster galaxies, indeed, can break but
farther from the center---in the regions that we do not consider
here, because there are no CO data for them. We also estimate the
stellar velocity dispersion by assuming the stellar disks to be
marginally stable. Generally, however, the stellar disks can have a
reserve of stability, i.e., they can be thicker and more tenuous:
the estimated gas volume density at the same surface density will
then be lower than that for the marginal case and, hence, the gas
pressure will be even lower. Thus, for overheated disks the
deviation from the Blitz--Rosolowsky dependence only increases.

\section{THE GAS COMPONENTS OF~THE~CLUSTER AND~FIELD GALAXIES}

Let us consider the possible causes of the differences noted above
for the field and cluster galaxies. An anomalous (higher than that
observed in the field galaxies) ratio of the molecular and atomic
hydrogen surface densities is possible, for example, when
overestimating the number of \H2 molecules (e.g., as a result of the
overestimation of the conversion factor~$\chi$) or due to the action
of the intergalactic gas ram pressure ``stripping'' the atomic disk
or if the environmental effects stimulate a more rapid transition of
the atomic gas to molecules. Let us analyze these possibilities.

\subsection{The Conversion Factor}

The difference in gas metallicity that appeared due to a different
star formation history (or, for example, because of the difference
in the initial mass function of massive stars) could be responsible
for the deviation of the conversion factor~$\chi$ from the standard
one for the cluster galaxies. For the Virgo cluster galaxies close
to the cluster center, an enhanced abundance of heavy elements, on
average, by 0.25--0.30 dex relative to the solar one was actually
pointed out (Shields et al. 1991; Skillman et al. 1996; Dors and
Copetti 2006). For most of the galaxies, the relationship between
metallicity and conversion factor $\chi\propto Z^{-1}$ proposed by
Boselli et al. (2002) is most likely valid; hence a twofold excess
of heavy elements can lead to an overestimation of $\chi$
approximately by a factor of 2 and, accordingly, to an
overestimation of the \H2 mass by a factor of 2. For some of the
galaxies, the dependence $\chi(Z)$ is probably stronger---for them,
the overestimation of the amount of molecular gas can reach an order
of magnitude (Magrini et al. 2011). However, even in this case, the
scatter on the dependence $\eta(P)$ can be explained by an enhanced
metallicity only for the galaxies of the first group, although it is
worth noting that the anomaly in $\eta$ for them is observed only on
the disk periphery, where, obviously, there are less heavy elements
and, hence, the overestimation of the amount of \H2 must be less
significant than that in the central regions. However, for the
galaxies of the second and third groups, the molecular gas fraction
can exceed $\eta$ for normal-metallicity galaxies by two orders of
magnitude (Figs. 1c, 1d) and this cannot be explained by a different
conversion factor.

It is argued in a number of works that the conversion factor must
depend on whether the molecular gas is in a diffuse state or in the
form of giant self-gravitating clouds. Nevertheless, this question
cannot yet be ultimately solved, although based on observations of
the Galaxy, Liszt et al. (2010) recently showed that the generally
not obvious but widespread assertion about universality of the
conversion factor for forms of \H2 is confirmed by observations.

\subsection{The Molecular Gas Fraction\\ with Respect to the Total Gas
Density}

\begin{figure*}
\includegraphics[width=16.0cm]{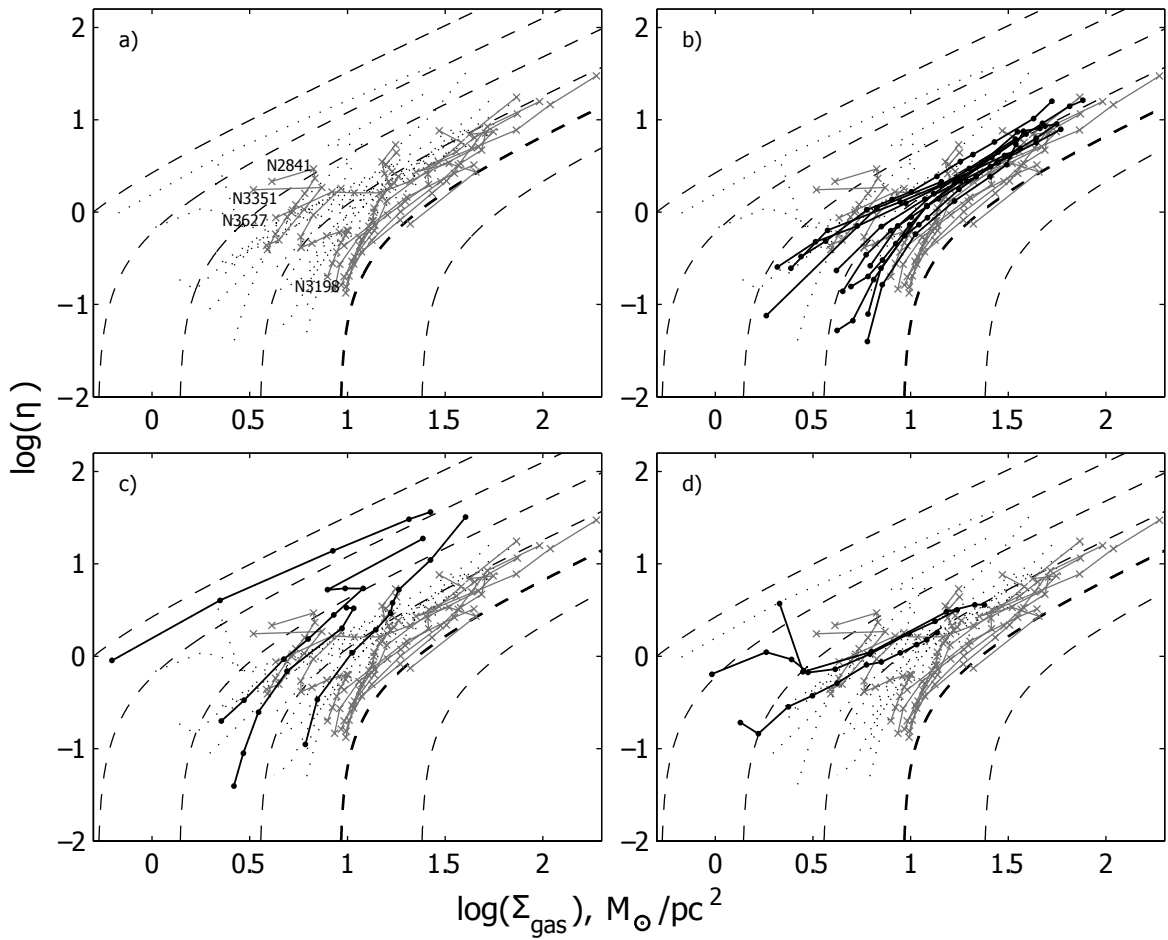}
\caption{Fraction of the molecular gas with respect to the atomic
one versus total gas surface density. The dashed lines indicate the
model dependences from Krumholz et al. (2009) for various
metallicities: from right to left, the lines correspond to
$log(Z/Z_{sun})$ equal to -0.5, 0, 0.5, 1, 1.5, and 2. The remaining
designations are the same as those in Fig. 1.}\label{fig2}
\end{figure*}

As we have already mentioned in the Introduction, Krumholz et al.
(2009) developed a model for the formation of molecular clouds whose
key parameters are the total gas surface density and the amount of
heavy elements in the interstellar medium. The higher the
metallicity, the more efficient the formation of \H2 molecules, and
the cold molecular clouds are shielded from the destructive action
of ultraviolet radiation as the thickness of the gas layer
increases. For solar metallicity at $\Sigma_{gas}<10$ {\MspcII}, the
optical depth is insufficient to preserve the \H2 molecules, and the
fraction $\eta$ increases sharply.

In Fig. 2, $\eta$ is plotted against the gas surface density
$\Sigma_{gas}=\Sigma_{HI}+\Sigma_{H_2}$ for our galaxies. The dashed
lines indicate the model dependences from Krumholz et al. (2009) for
various metallicities. Although the dependences are very similar in
pattern to $\eta(P)$, the correlation breaks down noticeably on the
plot of $\eta(\Sigma_{gas})$ for some of the normal spirals along
the radius (NGC 2841, NGC 3198, NGC 3351, and NGC 3627).
Nevertheless, on the whole, the $\eta(\Sigma_{gas})$ distributions
for the field galaxies and the Virgo cluster galaxies of the first
group, except the outer regions of the gas disks, agree with the
model curves, especially if we consider a slightly enhanced
metallicity for the cluster galaxies. However, for the galaxies of
the second and third groups within the framework of the model by
Krumholz et al., just as with the uncertainty of the conversion
factor mentioned above, we will have to assume the metallicity to be
more than an order of magnitude higher than the solar one, which is
inconsistent with the available observations. It also follows from
Fig. 2 that no distinct boundary surface density of about 10~\MspcII
below which the model by Krumholz et al. (2009) predicts a sharp
decrease in $\eta$ is observed for the cluster galaxies.

Thus, it seems obvious that to explain the ratio of the gas
components for the cluster galaxies, it is insufficient to consider
such internal gas characteristics as the metallicity and surface
density. The external factors related to the environmental effects
on the balance between the gas components should also be taken into
account.

\subsection{Peculiarities of the Radial\\ {\HI} and \H2 Distributions}

\begin{figure*}
\includegraphics[width=16.0cm]{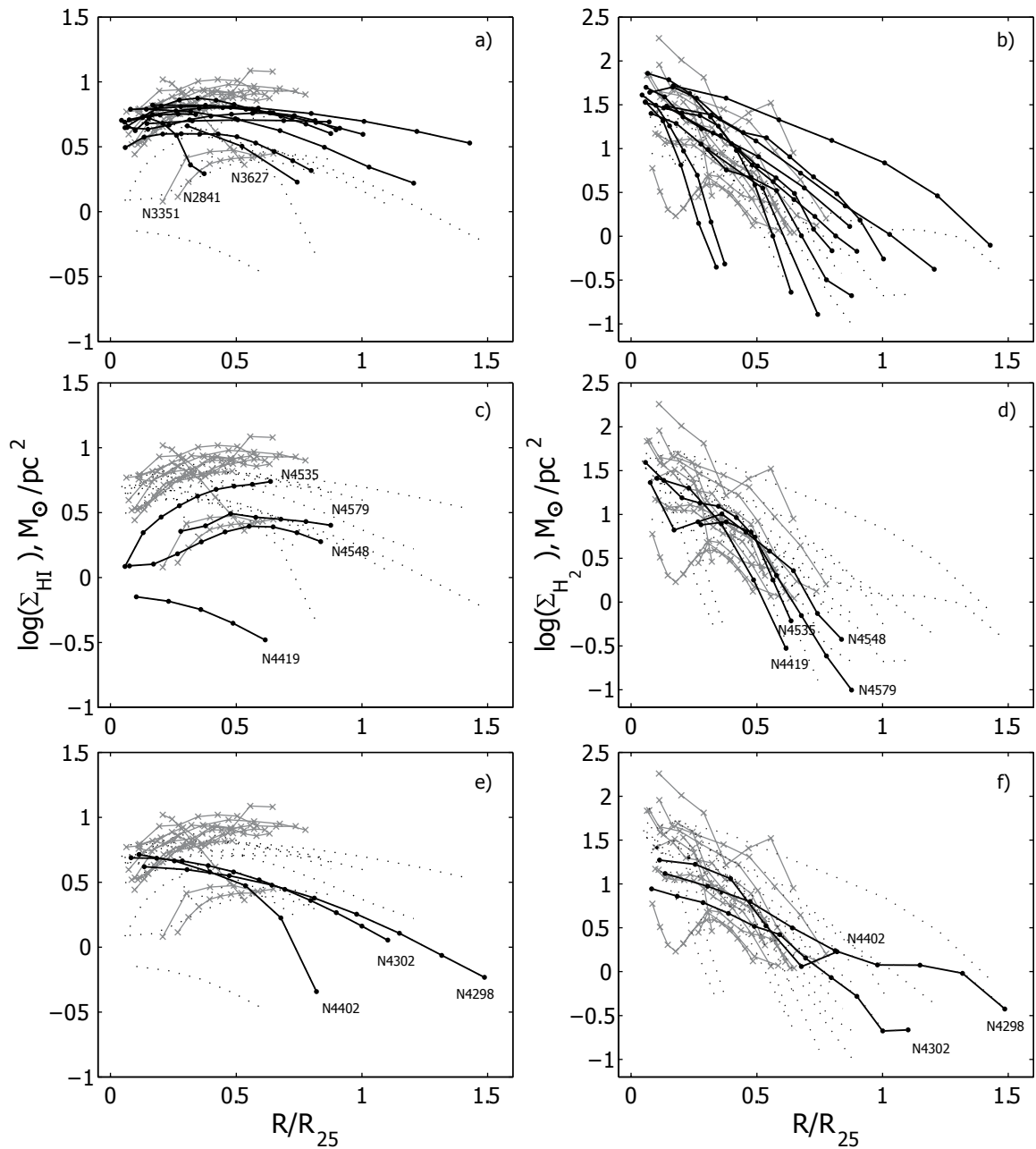}
\caption{Radial distributions of the atomic and molecular gas
surface densities: (a) and (b) the atomic and molecular surface
densities of the group I galaxies in comparison with those of the
field galaxies; (c) and (d) the same for the group II galaxies; (e)
and (f) for the group III galaxies. The designations are the same as
those in Fig. 1.}\label{fig3}
\end{figure*}

The distributions of the molecular and atomic gas surface densities
(Fig. 3) show that the most probable cause of the anomalously high
$\eta$ for some of the cluster galaxies under consideration is a low
atomic hydrogen abundance (at the center or on the periphery of the
disks) rather than a high \H2 abundance. A similar conclusion also
follows from the volume density distributions of the gas components
in the disk midplane that we derived. In comparison with the field
galaxies, an {\HI} deficiency is observed at all distances from the
center for the objects of the second group (Fig. 3c) and to a
greater or lesser extent on the periphery for the first and third
groups (Figs. 3a, 3e). It is worth noting that, just as for the
three field galaxies (NGC 2841, NGC 3351 è NGC 3627) that we
mentioned above, an {\HI} deficiency is observed at various
distances from the center due to the unusual behavior on the
dependences $\eta(P)$) and $\eta(\Sigma_{gas})$.

\begin{figure}
\includegraphics[width=8.0cm]{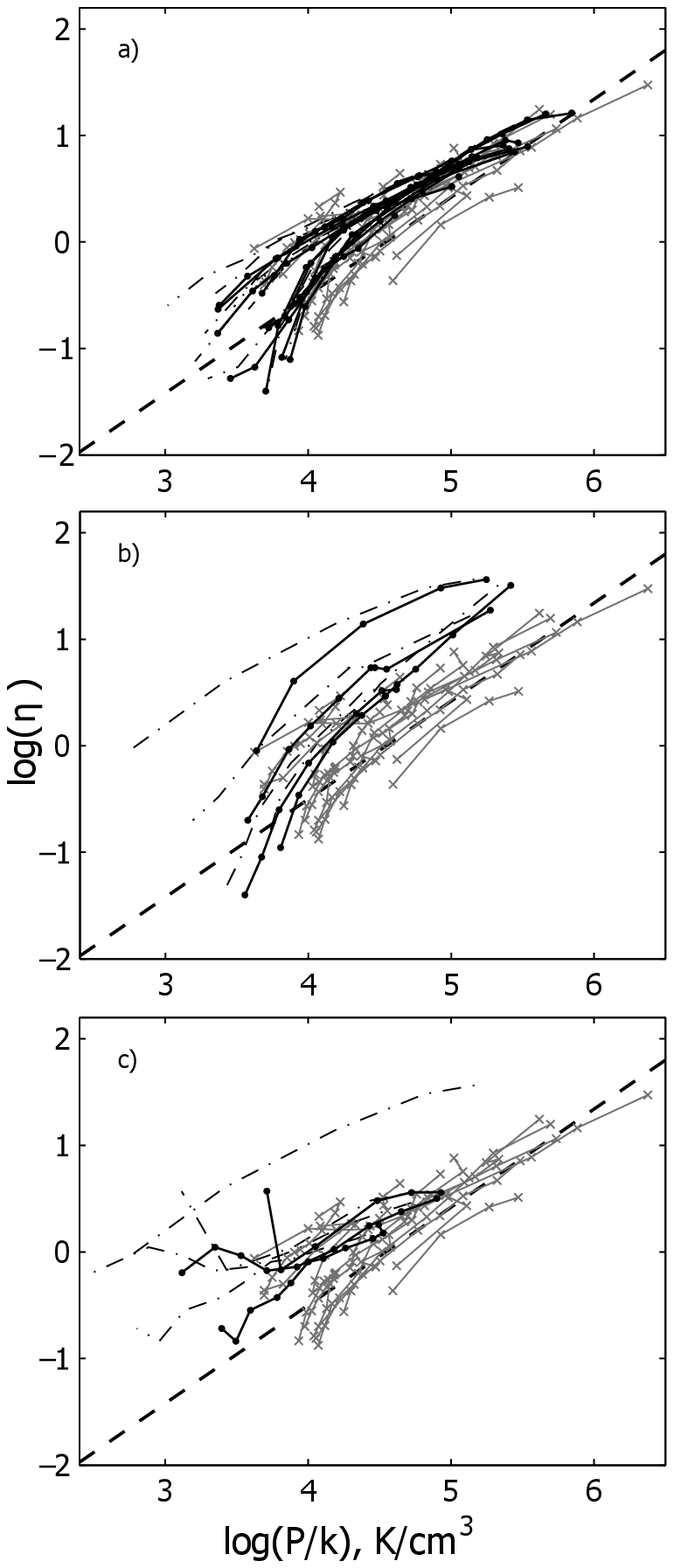}
\caption{Dependence of the molecular hydrogen fraction with respect
to atomic hydrogen on the turbulent gas pressure in the disk
midplane calculated for the model, average for a given morphological
type, $\Sigma_{HI}$ distribution taken from Cayatte et al. (1994).
For comparison, the dash--dotted line indicates the dependences for
the galaxies of the same subgroups but with the original, observed
atomic gas surface densities presented in~Fig.~1.}\label{fig4}
\end{figure}

The ram pressure of the intergalactic medium, which sweeps out the
diffuse gas, may be considered the most efficient {\HI} loss
mechanism (see Vollmer et al. 2001; Roediger 2009). Using a simple
procedure, it can be verified whether the observed {\HI} deficiency
is enough to explain the anomalous ratio of the gas components
$\eta$ for the cluster galaxies. Suppose that before the deficiency
of atomic hydrogen emerged in the galaxies, they had a radial {\HI}
distribution typical for field galaxies of a given type. We will use
the mean atomic hydrogen distributions for the galaxies of the
corresponding morphological types (see Cayatte et al. 1994) by
replacing the observed $\Sigma_{HI}(R)$ with them and leaving the
observed \H2 distribution unchanged. Subsequently, we will
recalculate the equilibrium gas pressure corresponding to these
``mixed'' data. The black  solid lines in Fig. 4 indicate the
dependences  $\eta(P)$ obtained in this way for the galaxies of
different subgroups. For clarity, the dash--dotted lines indicate
the initial dependences for the same groups of galaxies presented in
Fig. 1. As follows from Fig. 4a, about half of the galaxies from the
first group completely fell on the dependence typical of the field
galaxies, i.e., the observed amount of molecular gas in them agrees
well with that expected for the pressure $P$ existed before the loss
of \HI. For these galaxies, we clearly see a sharp decrease in
$\eta$ on the disk periphery. This was obtained in our calculations,
because the gas self--gravity and the dark halo effect were taken
into account. This peculiarity agrees well with the model by
Krumholz et al. (2009).

However, on the disk periphery for the rest of the galaxies from the
first and third groups and for the central regions of the galaxies
from the second group, the sweepout of {\HI} explains only partly
the high value of $\eta$ (Figs. 4b, 4c, and it becomes necessary to
assume an additional transition from {\HI} to \H2 due to some factor
that is most likely related to the environmental effects.

\section{CONCLUSIONS}

In this paper, we set the goal of investigating the
atomic-to-molecular gas ratio in galaxies that enter and do not
enter the Virgo cluster. The balance between the gas components of
the galaxies can change under the action of the intergalactic
medium, and the physics of one of the key star formation stages, the
transition from a warm atomic medium to cold molecular clouds, can
be traced by studying the peculiarities of the gas content.

To reveal the key characteristics of the interstellar medium
establishing a certain ratio of the {\HI} and \H2 components, we
compared two approaches: the dependence of the molecular gas
fraction $\eta$ predominantly on the equilibrium hydrostatic
turbulent pressure (Blitz and Rosolowsky 2006) or on the metallicity
and total gas surface density (Krumholz et al. 2008, 2009; McKee and
Krumholz 2010). The deviations from these correlations must be
indicative either of an imbalance between the components or
shortcomings of the equilibrium models. Both approaches
satisfactorily explain the observed ratios of {\HI} to \H2 for the
field galaxies, but there is a significant scatter in both cases for
the cluster galaxies (Figs. 1 and 2). On the whole, the correlation
$\eta(P)$ turned out to be slightly closer than
$\eta(\Sigma_{gas})$. This is particularly noticeable in some
segments of the anticorrelation in the $\eta(\Sigma_{gas})$
distribution that take place even for noninteracting galaxies.
Nevertheless, the significant scatter for the cluster members
compared to the field galaxies on both dependences requires
explanations. For convenience, we divided the sample into three
formal groups differing in behavior on the dependences under
discussion (see Section 2.2) by assuming that based on these
features, we can distinguish the various effects of the
intergalactic medium on the gas of the cluster galaxies.

In Section 3.3, we showed that the high molecular gas fraction for
the cluster galaxies could not be entirely explained by the
stripping of {\HI}. However, for most of the objects from the first
group, the scatter on the dependence $\eta(P)$ can be explained
almost entirely by a deficiency of atomic hydrogen at a normal \H2
abundance. The remaining excess of $\eta$ above the expected values
for several galaxies of the first group after allowance for the
{\HI} deficiency (Fig. 4a) can be explained by a lower conversion
factor $\chi$ for the cluster galaxies (Magrini et al. 2011). For
the galaxies of the second group, we should assume an excess \H2
abundance for a given $P$ at least in the central regions, because
even if the galaxy had no {\HI} deficiency, the deviation from the
dependence holding for the field galaxies cannot be removed (see
Fig. 4b). These galaxies are characterized by a significant
deficiency of both atomic and molecular gases. In contrast, the
objects of the third group have an unusually high \H2 fraction
primarily not at the center but on the periphery of the disks (Fig.
4c).

Thus, under the assumption made by Blitz and Rosolowsky (2006) about
the key importance of the interstellar gas pressure, we can entirely
explain the balance between the atomic and molecular components of
the gas medium for most of the field galaxies and, with allowance
made for the deficiency of \HI, for the first group of cluster
galaxies. However, some disregarded factor related to the evolution
of galaxies in a cluster and leading to growth of the molecular
fraction in the central regions and on the periphery of the disks,
respectively, obviously acts on the objects of the second and third
groups. We will devote the next paper to this subject matter.

The following question also remains open: why does not the molecular
gas density decrease, for example, due to star formation or the
destruction of molecular clouds by ultraviolet radiation after the
``removal'' of {\HI} from the galaxy or after the transition of the
bulk of {\HI} to \H2 (it does not matter through what processes)? A
possible way out of this situation would be a long life of molecular
clouds: to be more precise, the cloud destruction time must be
longer than the time elapsed after the decrease in the bulk of \HI.

According to the model by Krumholz et al.,molecular clouds at solar
metallicity can exist in considerable quantities only at a total gas
surface density greater than 10~\MspcII. We see from Fig. 2 that
this is not the case for many of the Virgo cluster galaxies, and
$\eta$ can be great even at a lower gas density. To explain the
behavior of most of these galaxies in terms of the model by Krumholz
et al. (2009), it should be assumed that the metallicity for them is
an order of magnitude higher than that for the field galaxies.
However, observations show a difference only by a factor of 2 or 3
(Shields et al. 1991; Skillman et al. 1996; Dors and Copetti 2006).
Nevertheless, we see good agreement with the model by Krumholz et
al. for the field galaxies and the first group of cluster galaxies.
In this model, the galaxies of the first group lie, on average,
slightly higher than the field galaxies on the $\eta(\Sigma_{gas})$
diagram by a value corresponding to a metallicity exceeding the
solar one approximately by 0.3 dex (Fig. 2b). However, the behavior
of the galaxies from the second and third groups on the dependence
$\eta(\Sigma_{gas})$ cannot be explained in such a way, and the
additional parameters reflecting the environmental effects on the
molecular gas fraction should be included in the description for
these objects.

\vspace{0.3cm}

\textbf{Acknowledgments.}

I wish to thank A.V. Zasov for his help in choosing the direction of
research, productive scientific discussions, and valuable remarks.
This work was in part supported by a grant from the President of
Russian Federation for the State Support of Young Russian PhDs (MK
73.211.2) and the grant RFBR 11-02-12247-ofi-m-2011.

\section{REFERENCES}

\renewcommand{\baselinestretch}{1}
\setlength{\parindent}{0cm}

1. O. V. Abramova and A. V. Zasov, Astron. Rep. \textbf{55}, 202
(2011).

2. E. F. Bell and R. S. de Jong, Astrophys. J. \textbf{550}, 212
(2001).

3. F. Bigiel, A. K. Leroy, F. Walter et al., Astrophys. J.
\textbf{730}, L13 (2011).

4. L. Blitz and E. Rosolowsky, Astrophys. J. \textbf{612}, L29
(2004).

5. L. Blitz and E. Rosolowsky, Astrophys. J. \textbf{650}, 933
(2006).

6. A. D. Bolatto, A. K. Leroy, E. Rosolowsky et al., Astrophys. J.
\textbf{686}, 948 (2008).

7. A. Boselli, J. Lequeux, and G. Gavazzi, Astron. Astrophys.
\textbf{384}, 33 (2002).

8. V. Cayatte, C. Kotanyi, C. Balkowski et al., Astron. J.
\textbf{107}, 1003 (1994).

9. E. J. Chung, M.-H. Rhee, H. Kim et al., Astrophys. J. Suppl. Ser.
\textbf{184}, 199 (2009a).

10. A. Chung, J. H. van Gorkom, J. D. P. Kenney et al., Astron. J.
\textbf{138}, 1741 (2009b).

11. O. Daigle, C. Carignan, P. Amram et al., Mon. Not. R. Astron.
Soc. \textbf{367}, 469 (2006).

12. N. A. Devereux, J. D. Kenney, and J. S. Young, Astron. J.
\textbf{103}, 784 (1992).

13. I. Dicaire, C. Carignan, P. Amram et al., Mon. Not. R. Astron.
Soc. \textbf{385}, 553 (2008).

14. O. L. Dors and M. V. F. Copetti, Astron. Astrophys. 452, 473
(2006).

15. B. G. Elmegreen, Astrophys. J. \textbf{338}, 178 (1989).

16. L. Ferrarese, W. L. Freedman, R. J. Hill et al., Astrophys. J.
\textbf{464}, 568 (1996).

17. M. Fumagalli, M. R. Krumholz, J. X. Prochaska et al., Astrophys.
J. \textbf{697}, 1811 (2009).

18. M. Fumagalli, M. R. Krumholz, and L. K. Hunt, Astrophys. J.
\textbf{722}, 919 (2010).

19. G. Gavazzi, A. Boselli, A. Donati et al., Astron. Astrophys.
\textbf{400}, 451 (2003).

20. P. Girichidis, C. Federrath, R. Banerjee et al., Mon. Not. R.
Astron. Soc. \textbf{413}, 2741 (2011).

21. S. C. O. Glover and P. C. Clark, arXiv: 1105.3073 (2011).

22. G. H. Heald, R. J. Rand, R. A. Benjamin et al., Astrophys. J.
\textbf{663}, 933 (2007).

23. A. V. Kasparova and A. V. Zasov, Astron. Lett. \textbf{34}, 152
(2008).

24. J. D. Kenney and J. S. Young, Astrophys. J. \textbf{301}, L13
(1986).

25. J. D. Kenney and J. S. Young, Astrophys. J. \textbf{344}, 171
(1989).

26. R. A. Koopmann, J. D. P. Kenney, and J. Young, Astrophys. J.
Suppl. Ser. \textbf{135}, 125 (2001).

27. M. R. Krumholz, C. F. McKee, and J. Tumlinson, Astrophys. J.
\textbf{689}, 865 (2008).

28. M. R. Krumholz, C. F. McKee, and J. Tumlinson, Astrophys. J.
\textbf{693}, 216 (2009).

29. A. K. Leroy, A. Bolatto, K. Gordon et al., Astrophys. J.
\textbf{737}, 12 (2011).

30. A. K. Leroy, F.Walter, E. Brinks et al., Astron. J.
\textbf{136}, 2782 (2008).

31. H. S. Liszt, J. Pety, and R. Lucas, Astron. Astrophys.
\textbf{518}, A45 (2010).

32. L. Magrini, S. Bianchi, E. Corbelli et al., Astron. Astrophys.
\textbf{535}, 13 (2011).

33. C. F. McKee and M. R. Krumholz, Astrophys. J. \textbf{709}, 308
(2010).

34. C. A. Narayan and C. J. Jog, Astron. Astrophys. \textbf{394}, 89
(2002).

35. K. Nishiyama, N. Nakai, and N. Kuno, Publ. Astron. Soc. Jpn.
\textbf{53}, 757 (2001).

36. E. Roediger, Astron. Nachr. \textbf{330}, 888 (2009).

37. E. Rosolowsky and L. Blitz, Astrophys. J. \textbf{623}, 826
(2005).

38. V. C. Rubin, A. H. Waterman, and J. D. P. Kenney, Astron. J.
\textbf{118}, 236 (1999).

39. G. A. Shields, E. D. Skillman, and R. C. Kennicutt, Astrophys.
J. \textbf{371}, 82 (1991).

40. E. D. Skillman, R. C. Kennicutt, G. A. Shields et al.,
Astrophys. J. \textbf{462}, 147 (1996).

41. Y. Sofue and V. Rubin, Ann. Rev. Astron. Astrophys. \textbf{39},
137 (2001).

42. B. Vollmer, V. Cayatte, C. Balkowski et al., Astrophys. J.
\textbf{561}, 708 (2001).

\renewcommand{\baselinestretch}{1}

\vspace{0.5cm} {\textit{Translated by V. Astakhov}
\end{document}